\documentclass[runningheads]{llncs}
\usepackage[T1]{fontenc}
\usepackage{graphicx}
\usepackage{amsmath,amsfonts}
\usepackage{tabularx}
\usepackage{booktabs}
\usepackage{enumitem}
\usepackage{caption}
\usepackage{subcaption}
\usepackage{xspace}
\usepackage{cite}
\usepackage{adjustbox}
\usepackage{multirow}

\usepackage{hyperref,color}

\def\fullname{Triumvir\xspace}

\def\encod{EnCoD\xspace}

\DeclareMathOperator*{\argmax}{arg\;max}
\DeclareMathOperator*{\argmin}{arg\;min}

\title{When Entropy Is Not Enough: Multi-Modal Classification of Encrypted and Compressed Data Fragments}

\begin{document}

\author{Fabio De Gaspari\inst{1}\orcidID{0000-0001-9718-1044} \and
Dorjan Hitaj\inst{1}\orcidID{0000-0001-5686-3831} \and
Samuele Salaris\inst{1} \and
Luigi V. Mancini\inst{1}\orcidID{0000-0003-4859-2191}}
\authorrunning{De Gaspari et al.}
%
\institute{
Dipartimento di Informatica, Sapienza University of Rome, Rome, Italy\\
\email{\{surname\}@di.uniroma1.it}, salaris.2083011@studenti.uniroma1.it} 
\maketitle             

\begin{abstract}
Reliable identification of encrypted data fragments is essential in cybersecurity, with applications to ransomware detection, digital forensics, and large-scale data analysis. Distinguishing encrypted from compressed fragments is particularly challenging, as short fragments lack structural data and exhibit low statistical redundancy. Traditional statistical methods based on byte-level distributions show limited effectiveness on this task. Recent machine learning approaches improve performance by learning subtle patterns from raw bytes, but predominantly rely on single-modal representations, implicitly assuming that a single view of the data is sufficient for accurate classification.

This paper shows that this assumption becomes a fundamental limitation in low-information settings, when only small fragments of data are available (512--2048 Bytes). We propose \textit{\fullname}, a multi-modal, uncertainty-aware ensemble architecture that integrates statistical, sequential, and spatial representations of raw byte fragments. Extensive experimental analysis demonstrates that \textit{\fullname} consistently outperforms state-of-the-art methods with gains of up to $+4.5pp$ in binary and $+6.4pp$ in multiclass classification. Ablation studies confirm that combining modalities is critical, yielding improvements of up to $+5pp$ over partial configurations.

\end{abstract}
\section{Introduction}\label{sec:introduction}

Digital forensics, incident response, and network security operations increasingly rely on automated analysis of raw byte fragments: data segments stripped of file-system metadata, application context, and structural headers. In such settings, the ability to identify the type of file fragments is a fundamental capability for a broad class of security tasks~\cite{de2022reliable,shapira2021flowpic,casino2019hedge}. As modern applications heavily rely on compression, distinguishing between encrypted and compressed raw data fragments is a particularly critical challenge, with direct implications for ransomware detection, memory forensics, network traffic analysis, and data exfiltration detection~\cite{hitaj2025minerva,continella2016shieldfs,de2017botnet,sabir2021machine}. However, this task is inherently challenging. Compression algorithms are specifically engineered to remove statistical redundancy, producing compressed data that often exhibit statistical properties that closely resemble those of encrypted data.
Moreover, structural elements of compressed data are often lost in small data fragments, further complicating the identification. Under these conditions, both encrypted and compressed data fragments exhibit high entropy and near-uniform byte distributions, rendering traditional statistical techniques largely ineffective~\cite{de2022reliable}. As fragment size decreases, the statistical differences between the two classes shrink further, making reliable discrimination increasingly difficult.

Existing approaches to this problem can be categorized in two broad classes: \textit{Context-dependent}, and \textit{context-agnostic}. Contex-dependent approaches leverage auxiliary information such as network flow dynamics~\cite{saleh2023combining,yan2025high}, process behavior~\cite{continella2016shieldfs}, or system-level correlations~\cite{hitaj2025minerva}, achieving high performance within specific domains. While effective, their reliance on external context limits their applicability and generality. In contrast, context-agnostic techniques operate directly on raw data fragments. Classical statistical techniques such as Shannon entropy and chi-squared test~\cite{casino2019hedge} are inherently context-agnostic, but perform poorly in the small-fragment-size regime~\cite{de2022reliable}. On short fragments, the statistical distance between encrypted and compressed data is small and highly variable, leading to high misclassification rates at sub-4KB granularities~\cite{de2020encod}. Machine learning based approaches currently represent the state-of-the-art, as they learn subtle discriminative patterns from raw data fragments that allow effective type classification. However, current leading ML approaches predominantly rely on single-modal representations such as histogram-based features~\cite{de2022reliable}, implicitly assuming that that one view of the data is sufficient for accurate classification.

This paper argues that this assumption is a fundamental limitation. While encrypted and compressed data may appear indistinguishable under simple first-order statistical analysis, they differ in sequential structure and local regularities introduced by compression algorithms. To address this limitation, we propose \textit{\fullname}, a multi-modal uncertainty-aware ensemble architecture for context-free data fragment classification. \fullname leverages three complementary signal spaces, captured by three different expert models: 1) sequential space, capturing sequential patterns introduced by back-references and repetition coding in compression algorithms; 2) spatial space, capturing local patterns and regularities in the byte layout; and 3) statistical space, capturing statistical properties of the byte-level distribution. The predictions of the three expert model are combined through a learned gate network that weights their individual contributions. We further enhance robustness by integrating per-expert temperature scaling and margin-based confidence, providing an explicit confidence signal to the gate network. Our extensive experimental analysis shows that the proposed architecture consistently outperforms existing approaches across a wide range of different file types and fragment sizes, particularly in the challenging small-fragment regime where traditional techniques fail.

In summary, we make the following contributions:
\begin{itemize}
    \item We identify the reliance on single-modal representation as a key limitation of current context-free approaches and empirically confirm that encrypted and compressed data fragments become indistinguishable at short fragment lengths under simple first-order byte analysis.
    \item We propose \fullname, a multi-modal uncertainty-aware ensemble architecture that captures complementary sequential, spatial, and statistical signals present in compressed data to accurately classify file type fragments. \fullname integrates a confidence-aware learned gate network that combines calibrated expert predictions, implicitly adapting the relative contribution of each expert based on the specific input.
    \item We perform an extensive experimental analysis showing that \fullname consistently outperforms multiple baseline approaches across a wide range of file types and fragment sizes.
\end{itemize}

The rest of this paper is structured as follows: Section~\ref{sec:bck_related_wk} provides the necessary background knowledge and reviews existing approaches to the problem. Section~\ref{sec:threat_model} describes the problem setting and Section~\ref{sec:our_approach} presents \fullname, a novel approach to the problem. 
Section~\ref{sec:experimental_setup} provides details about the experimental setup on which we evaluate \fullname.
Section~\ref{sec:evaluation} evaluates the performance of the considered approach, discussing the strengths and limitations, and Section~\ref{sec:conclusion} concludes the paper.

\section{Related Works}\label{sec:bck_related_wk}

Determining the format of a data object, such as a file in storage or an HTTP payload, is a common task. Under normal conditions, this can be achieved using metadata or by directly parsing the object. However, the problem becomes significantly more challenging when metadata is absent and the data is corrupted or incomplete.

\leavevmode \\
\textbf{Entropy-based encryption detection.}
Use of entropy estimation to detect encrypted content is common in ransomware detection. Proposals such as RWGuard~\cite{mehnaz_rwguard}, UNVEIL~\cite{kirda_unveil}, Redemption~\cite{kirda_redemption}, Minerva~\cite{hitaj2025minerva} and ShieldFS~\cite{continella_shieldfs:_2016} use entropy of written content either directly as a feature, or as part of feature calculation. It should be noted that none of these detectors use entropy as the sole feature for detection. In the realm of digital forensics, entropy estimation has been used to determine the type of unknown disk data fragments. One of the most complete approaches is that of Conti et al.~\cite{conti_automated_2010}. 

Entropy estimation has also been applied to the real-time analysis of network traffic. Zhang et al. proposed an entropy-based classifier for the identification of botnet traffic~\cite{zhang_detecting_2013}. All these approaches also suffer from the limitations of using high entropy as a fingerprint of encryption.
MovieStealer~\cite{wang_steal_2013} aims at identifying encrypted and decrypted-but-compressed media buffers in order to break DRM. It uses an entropy test to single out encrypted and compressed buffers from other data, and the $\chi^2$-test to distinguish them. It requires 800KB of data to reliably identify random data, which is far beyond the fragment size in the scenarios that we consider.
\leavevmode \\
\textbf{Non-entropy-based approaches.}
HEDGE, proposed by Casino et al.~\cite{casino_hedge_2019}, combines the $\chi^2$ test with a subset of the NIST suite~\cite{rukhin_statistical_2010} to distinguish between encrypted and compressed network traffic. 
A key limitation of this class of methods is their relatively low accuracy, particularly when operating on small data blocks. More broadly, statistical randomness–based approaches (e.g.,~\cite{lipmaa_data_2017,choudhury_empirical_2020}) lack the capacity to differentiate between distinct types of compressed archives, restricting their applicability in more fine-grained classification tasks.
Mbol et al.~\cite{foresti_efficient_2016} explore the use of Kullback–Leibler divergence to distinguish encrypted files from JPEG images. While their results demonstrate the potential of divergence-based metrics, the analysis is limited in scope, focusing solely on a single file format comparison. Moreover, their approach assumes access to relatively large contiguous data blocks (128–512 KB) extracted from the beginning of each file. In practical scenarios—particularly in digital forensics and network traffic analysis—such uninterrupted segments are rarely available, which constrains the method’s real-world applicability.
\leavevmode \\
\textbf{Learning based approaches.}
The abovementioned work suggest that tests based on byte-value distribution, such as $\chi^2$, can distinguish some encrypted and compressed content, but have accuracy issues. Such tests collapse the entire distribution to a single scalar value, losing information concerning the shape of the distribution. As such, research naturally was drawn towards ML to improve such results given the fact that ML solutions can consider the entire discrete distribution (modeled as a feature vector), and can learn to recognize complex distributions~\cite{Lee0MRA17}. Initially, Ameeno et al.~\cite{ameeno_using_2019} showed promising preliminary results, however the analysis is limited in scope: it only attempts to distinguish zip archives from rc4-encrypted data, and considers whole files (not fragments). 
To the best of our knowledge, the first learning-based approach is Encod~\cite{de2020encod}, which uses deep neural networks trained on byte-value histograms to distinguish compressed from encrypted data. This method represents the current state-of-the-art, achieving accuracies of ranging from 0.82 on 512-byte fragments and up to 0.92 on 8KB fragments. However, such approaches rely on single-modal representations~\cite{de2020encod}, implicitly assuming that a single view of the data suffices for accurate classification. 
In practice, this assumption limits performance, particularly on small fragments where discriminative signals are weaker. This limitation stems from the fact that, although encrypted and compressed data appear similar under first-order statistical analysis, they differ in sequential structure and local regularities introduced by compression algorithms. Addressing this gap is the central motivation behind \fullname.
\section{Problem Setting and Assumptions}\label{sec:threat_model}
\label{sec:problem_setting}

We consider the problem of classifying data fragments solely based on their raw byte content, in the absence of contextual information or structural metadata. This problem has direct implications for a broad class of security tasks, including forensics analysis, ransomware detection, and network traffic analysis, where only partial observations of data are available. The goal is to learn a context-free classifier that can perform both binary classification of encrypted and compressed fragments, as well as multi-class file type classification, using only byte-level features.

More formally, let $x = (x_1, x_2, ..., x_n)$ denote a byte-level fragment of size $n$ bytes, with $x_i \in [0,255] \; \forall i$. The goal is to learn two classification functions:

\begin{center}
    $f_b(x) \rightarrow y_b$; \; $f_m(x) \rightarrow y_m$ 
\end{center}

where $y_b$ and $y_m$ correspond to binary and multi-class file type labels, respectively. The objective is to learn $f(\cdot)$ only from the byte-level fragment $x$.

\paragraph{\textbf{Data Fragments.}}
The input consists of byte sequences (\textit{fragments}) of varying length extracted from larger data sources. Fragments originate from arbitrary offsets within files or streams, and therefore may lack headers and format-specific structures. Consequently, classification relies exclusively on intrinsic properties of the fragment byte sequence. Fragments belong to one of two primary classes: \textit{encrypted} or \textit{compressed}. Compressed fragments may originate from a variety of different file types, compression schemes, and compression parameters used by their respective application programs. 

\paragraph{\textbf{Assumptions and Constraints.}}
We assume a context-agnostic setting where no auxiliary information is available. The classifier does not have access to file system metadata, protocol information, process context, or relationships between different fragments. Each fragment is analyzed and classified independently. We further assume that fragments are of sufficient size to retain residual signals useful for characterization, typically ranging from a few hundred bytes to a few kilobytes in size~\cite{de2022reliable}.

\section{\fullname}\label{sec:our_approach}
\label{sec:methodology}

We propose \fullname, a multi-modal 
ensemble architecture for data fragment classification that integrates heterogeneous representations of raw byte sequences. The primary idea is that encrypted and compressed fragments carry discriminative signals in multiple complementary feature spaces: sequential byte structure, spatial layout, and byte-frequency statistics. While no feature space is individually sufficient, their combination provides a strong discriminative signal that enables effective classification. For each signal space, \fullname leverages a specialized expert model with an appropriate inductive bias for the specific type of features. Expert outputs are aggregated through a learned gate network that takes as input the experts' calibrated logits alongside per-expert confidence margins, implicitly weighting expert contributions based on an approximation of their prediction uncertainty.

\begin{figure}[t]
    \centering
    \includegraphics[width=0.9\linewidth]{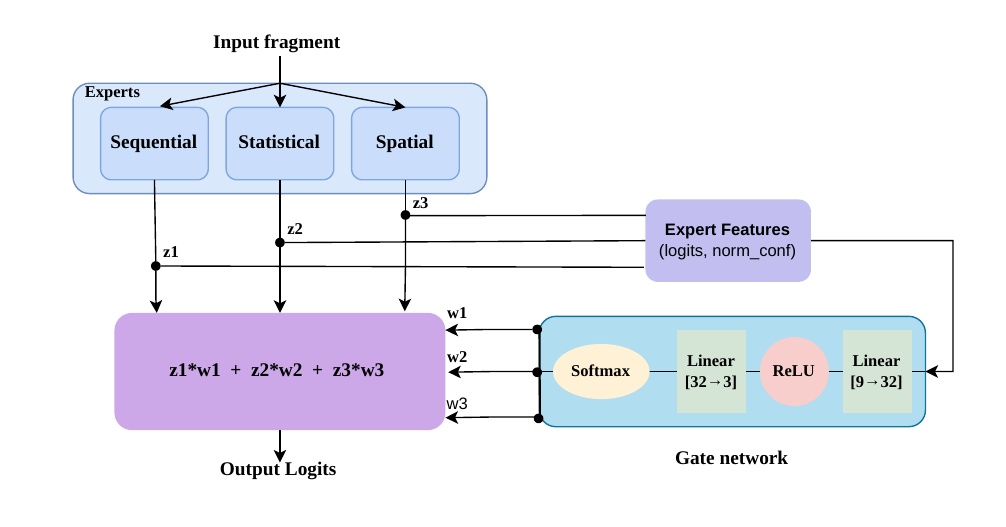}
    \caption{\fullname high-level overview.}
    \label{fig:arch}
\end{figure}

Figure~\ref{fig:arch} provides a high-level overview of \fullname architecture. 
In the following sections, we present expert architecture, logit calibration and confidence scoring, and gating.

\subsection{Expert Models}
\label{sec:expert models}

\paragraph{\textbf{Sequential Expert.}}
The sequential expert is a transformer encoder~\cite{vaswani2017attention} operating over the raw byte sequence of the input fragments. Fragment bytes are embedded and processed through multi-head self-attention layers, capturing long-range dependencies and repetition structure across the full fragment. This choice is motivated by the observation that compression algorithms, such as DEFLATE~\cite{deflate}, introduce back-references and repetition coding that create structured sequential patterns in the byte stream. The transformer attention mechanism is well-suited to capture these long-range correlations, which are not present in encrypted data. 

\paragraph{\textbf{Spatial Expert.}}
The structural expert is a Convolutional Neural Network (CNN) operating over a two-dimensional representation of the input fragments. This representation is obtained by reshaping the byte sequence to a matrix form and treating individual byte values as pixel intensities. This representation is well-suited to highlight spatial properties such as local patterns regularities in the byte sequence that are harder to identify in sequential and statistical representations. Compression algorithms produce characteristic block structures and utilize specific header markers that appear as spatial patterns in this representation space~\cite{deflate}. By contrast, encrypted data typically produces flat, noise-like matrices without any regular patterns. The convolutional filters of CNNs are well-suited to capture this type of local spatial artifacts.

\paragraph{\textbf{Statistical Expert.}}
The statistical expert is a multilayer perceptron (MLP) operating on a 256-dimensional byte-frequency histogram that is computed over the input fragment. Each element of the histogram feature vector represents the frequency of the $i$-th byte value, normalized by the fragment size. This representation directly captures the marginal distribution of each byte value within the data fragment, and is the primary representation space used by existing state-of-the-art approaches~\cite{de2022reliable}.

\subsection{Confidence Calibration}
\label{sec:confidence}
Raw neural network outputs are known to be poorly calibrated, with prediction scores not reliably reflecting true classification accuracy~\cite{moon2020confidence}. In \fullname's architecture, miscalibrated expert outputs may cause the gate network to incorrectly weight expert contributions. To address this, we apply per-expert temperature scaling~\cite{guo2017calibration} to calibrate output logits before gating. Formally, let $\phi(x) \in \mathbb{R}^{|Y|}$ be the logit output of an expert model on input $x \in \mathcal{X}$, and $\phi^{(k)}(x)$ its $k$-th component. We learn a temperature scaling parameter $T \in \mathbb{R}^+$ per expert by optimizing the negative log likelihood over a held-out validation set $\mathcal{D}_{\textbf{val}}$:

\begin{align}
    \argmin_{T\in \mathbb{R}^+} \; -\sum_{(x_i, y_i) \in \mathcal{D_{\text{val}}}} \log p(y_i | x_i, T)
    \\
    p(y \mid x, T) = \frac{e^{\phi^{(y)}(x)/T}}{\sum_{j \in \mathcal{Y}} e^{\phi^{(j)}(x)/T}}
\end{align}

where $y_i \in \mathcal{Y}$ is the ground-truth label for sample $x_i$. Temperature scaling leaves the relative ranking of predictions unchanged, while adjusting the spread of the output distributions. 
In addition, we compute a per-expert confidence margin score to provide an explicit uncertainty signal to the gate network. The confidence margin for expert $e$ on input $x$ is defined as the difference between the top-2 calibrated logits values:

\begin{align}
    m_e(x) = \hat{\phi}^{({k_1})}(x) - \hat{\phi}^{({k_2})}(x)
\end{align}

where $k_1 = \argmax_k \hat{\phi}^k(x)$, $k_2 = \argmax_{k \neq k_1} \hat{\phi}^k(x)$, and $\hat{\phi}$ are the temperature scaled logits. Confidence margins are computed directly on the calibrated logits rather than softmax probabilities, retaining raw magnitude information that would otherwise be suppressed~\cite{liang2025revisiting}.

\subsection{Gate Network}
The gate network is a shallow multilayer perceptron $g(\cdot)$ that weights expert outputs to produce the final classification. It takes as input the concatenation of the three experts' calibrated logit vectors and their corresponding confidence margins, and outputs a set of normalized weights for each expert: 

\begin{equation}
\begin{gathered}
f(\mathbf{u}) = softmax(g(\mathbf{u}))
\\
\mathbf{u} = \left[ \hat{\phi}_T(x)\mathbin\Vert \hat{\phi}_S(x)\mathbin\Vert \hat{\phi}_C(x)\mathbin\Vert m_T(x)\mathbin\Vert m_S(x)\mathbin\Vert m_C(x) \right]
\end{gathered}
\end{equation}

where $\hat{\phi}_T$, $\hat{\phi}_S$, $\hat{\phi}_C$ denote the temperature-scaled logits of the sequential, spatial, and statistical experts, and $m_T(x)$, $m_S(x)4$, $m_C(x)$ their confidence margins. The final classification is then given by the weighted sum of the experts' logits $\hat{\phi}$:

\begin{equation}
\hat{y} = \argmax f(\mathbf{u})^T \hat{\phi}(x);
\end{equation}

By conditioning on both logits and confidence margins, the gate network can weight expert contributions based on per-sample prediction confidence. The gate network is trained using cross-entropy loss, with expert weights and temperature parameters fixed.

\section{Experimental Setup}\label{sec:experimental_setup}

\begin{table}[t]
\setlength{\tabcolsep}{3.9pt}
\centering
\scriptsize
\caption{Summary of model architectures and training configuration. BN: BatchNorm; FC: Fully Connected; frz: frozen layers; '@': dataset size-specific  setting.}
\label{tab:impl_details}
\begin{tabular}{lcccc}
\toprule
\textbf{Component} & \textbf{Task} & \textbf{Architecture} & \textbf{\# Samples} & \textbf{LR} \\
\midrule
ByT5 & Pre-train & ByT5-small (12L) & 280K & $1\mathrm{e}{-4}$ \\
ByT5 & Binary & 12L (9 frz) + 2 FC & 20K & $1\mathrm{e}{-5}$; $5\mathrm{e}{-5}$ \\
ByT5 & Multi & 12L (9 frz) + 2 FC & 100K(20K@4;8KB) & $1\mathrm{e}{-5}$; $5\mathrm{e}{-5}$ \\
CNN & Bin/Multi & 3×(Conv+ReLU+BN+Pool) + 2 FC & 6M & $1\mathrm{e}{-5}$ \\
MLP & Binary & 3 FC (ReLU) & 6M & $1\mathrm{e}{-3}$ \\
MLP & Multi & 5 FC (SELU) & 6M & $1\mathrm{e}{-3}$ \\
Gate & Binary & 2 FC & 20K & $3\mathrm{e}{-4}$ \\
Gate & Multi & 2 FC & 128K (64K@8KB) & $3\mathrm{e}{-4}$ \\
\bottomrule
\end{tabular}
\end{table}

\subsection{Baseline Approaches and Dataset}
We compare \fullname against three baseline approaches: EnCoD~\cite{de2022reliable}, HEDGE~\cite{casino2019hedge}, and NIST test suite~\cite{rukhin_statistical_2010}. EnCoD represents the current state of the art ML approach for compressed and encrypted fragment classification. It uses an MLP operating on a 256-dimensional byte-frequency histogram and supports both binary and multiclass classification. HEDGE is a classical statistical approach originally proposed for classification of encrypted and compressed network packets. It combines multiple statistical tests, including Shannon entropy and chi-squared test, into a binary classification pipeline. The NIST statistical test suite is a standard battery of randomness tests developed to evaluate cryptographic random number generators. In line with previous works, we use the full suite to each fragment and assign the final class based majority vote. HEDGE and the NIST suite are strictly binary classifiers and are not applicable to multiclass classification. Classification performance is taken from~\cite{de2022reliable}.

For a fair comparison with the baseline approaches, we use the \encod~\cite{de2020encod} dataset with fragment sizes of 512, 1024, 2048, 4096, and 8192 bytes. The dataset comprises 400 million encrypted and compressed fragments spanning 17 different data formats, covering all major content types and encrypted data. Due to space constraints, we defer additional details to Appendix~\ref{sec:dataset}.

\subsection{Implementation Details}
We use ByT5 small~\cite{xue2022byt5} as the sequential expert model and fine-tune it in an unsupervised manner on a mixed corpus of raw file fragments. The spatial expert is implemented as a CNN with operating on a 2-dimensional reshaping of the input sequence. The statistical expert is an MLP with ReLU and SELU activations, taking as input the normalized byte histogram. 

Each expert is trained independently for both binary and multiclass classification tasks. The gate network is a fully-connected neural network which takes as input the temperature-scaled logits and margin based confidence scores of each expert (see Section~\ref{sec:confidence}). During gate training, the experts are frozen. Temperature scaling is applied and tuned independently for each expert using a held-out validation set. The gate network is trained independently for binary and multi-class classification using the corresponding set of experts. All models are trained with the AdamW optimizer with early stopping. Table~\ref{tab:impl_details} summarizes key implementation details.

\section{Evaluation}
\label{sec:evaluation}
We evaluate \fullname across four dimensions: statistical characterization of the problem (Section~\ref{sec:stat_analysis}), binary classification performance (Section~\ref{sec:binary_analysis}), multiclass classification performance (Section~\ref{sec:multiclass_analysis}), and ablation study (Section~\ref{sec:ablation_analysis}). All experiments are conducted on the~\cite{de2022reliable} dataset described in Section~\ref{sec:dataset}, across fragment sizes ranging from 512B to 8KB.

\begin{figure}[t]
  \begin{subfigure}[t]{\columnwidth}
    \includegraphics[width=0.9\textwidth]{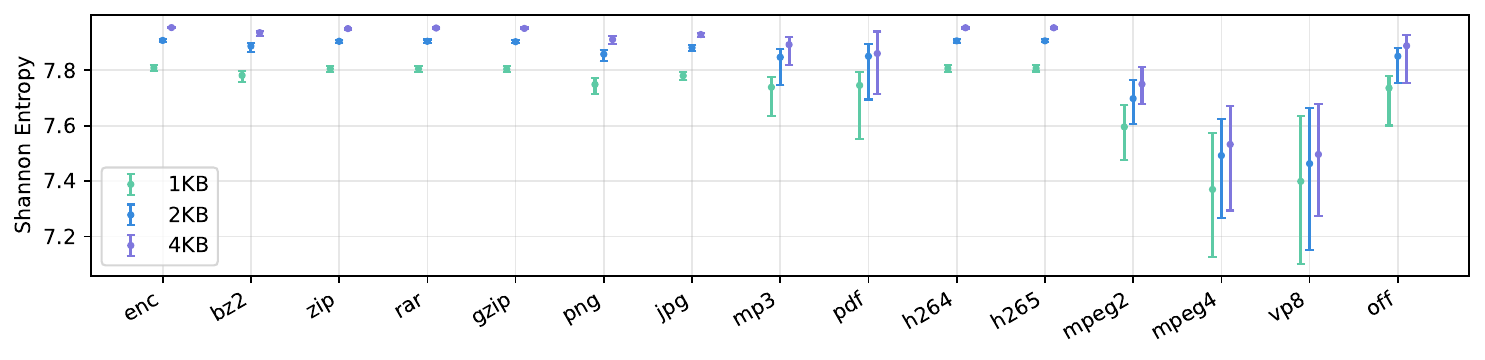}
    \caption{Entropy distribution.}
    \label{fig:entropy_dist}
  \end{subfigure}
  
  \begin{subfigure}[t]{\columnwidth}
    \includegraphics[width=0.9\textwidth]{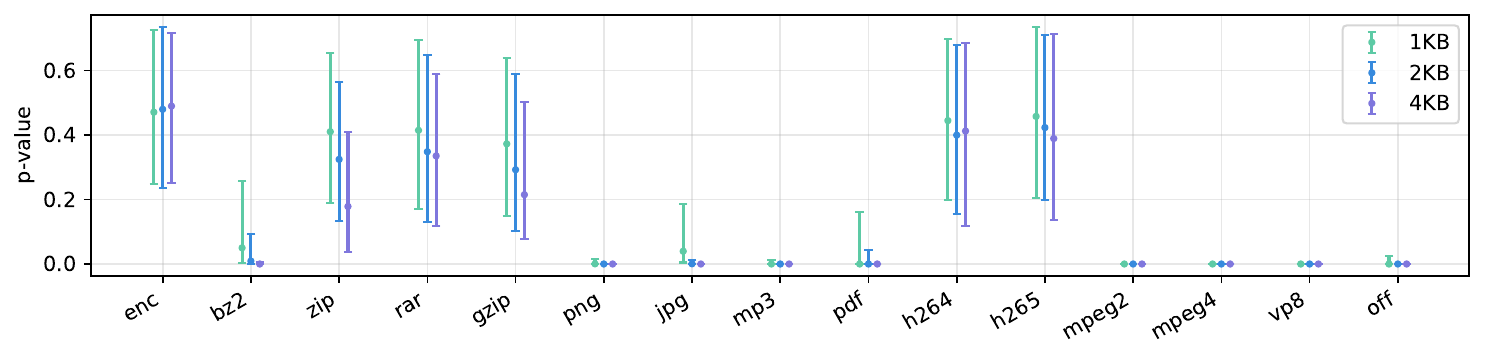}
    \caption{$\chi^2$ p-value distribution.}
    \label{fig:chi_distr}
  \end{subfigure}
  \caption{Shannon entropy and $\chi^2$ p-value for encrypted and compressed types. The dot is the median; error bars indicate 1st and 3rd quartiles.}
  \label{fig:fragment_dist}
\end{figure}

\subsection{Statistical Analysis}
\label{sec:stat_analysis}
We first characterize the statistical signal collapse under small fragment sizes by measuring the Shannon entropy and $\chi^2$ p-value overlap between encrypted and compressed file fragments. This analysis motivates the need for multi-modal classification and provides a quantitative baseline for the difficulty of the problem. Figure~\ref{fig:fragment_dist} reports the two distributions. Figure~\ref{fig:entropy_dist} reveals that entropy values converge toward the maximum across almost all file types, with large variance especially in smaller fragment sizes. All generic compression formats exhibit median entropy values at or above $7.8$ and are indistinguishable from encrypted data under this metric. Notably, as fragment size increases, most compressed formats continue to converge toward the entropy of encrypted data, highlighting that, even at larger sizes, entropy does not provide discriminative signal.

The $\chi^2$ p-value plot (Figure~\ref{fig:chi_distr}) reveals a more nuanced picture. Under the null hypothesis of uniformity, encrypted fragments produce p-values with medians around $\sim0.49$ and wide quartile ranges, consistent with near-uniform distribution. Most generic compression archives show similar p-value distributions, especially on lower fragment sizes, and substantial overlap with encrypted data. These results confirm that generic compressed formats are statistically similar to encryption under the $\chi^2$ metric. In contrast, most structured compressed formats such as png, jpg, mpeg, and office all exhibit very low p-values, indicating strong deviations from the uniform distribution. This result indicates that these formats retain generally retain sufficient byte-frequency structure to separate them from encryption under the $\chi^2$ metric. h264 and h265 are outliers among video formats, showing p-values in line with encryption and generic compression file types. As we will see, the inherent difficulty distinguishing these two formats will be later confirmed by our binary analysis.

\paragraph{Implications.} Taken together, these results indicate two distinct levels of statistical indistinguishability. A first group of structured compressed formats retains sufficient byte-frequency information to be partially separable from encrypted data (png, jpg, mp3, pdf, mpeg, vp8, office). A second, harder group, generic compression and h264/5, substantially overlaps the encrypted distribution under both entropy and $\chi^2$ metrics. As we will show in the following sections, it is exactly on this second group that \fullname achieves the larges average gains over baseline approaches (with a single exception, h265). This confirms that the sequential and spatial experts capture complementary signals that are fundamentally inaccessible to statistical-only approaches.

\subsection{Binary Classification Analysis}
\label{sec:binary_analysis}
\begin{table}[t]
\centering
\scriptsize
\caption{Per-file type binary accuracy on 2KB size. Best results underlined.}
\label{tab:2k_binary}
\resizebox{\textwidth}{!}{%
\begin{tabular}{l|cccccccccccccc}
\toprule
\textbf{Method} & \textbf{bz2} & \textbf{zip} & \textbf{rar} & \textbf{gzip} & \textbf{png} & \textbf{jpeg} & \textbf{mp3} & \textbf{pdf} & \textbf{h264} & \textbf{h265} & \textbf{mpeg2} & \textbf{mpeg4} & \textbf{vp8} & \textbf{office} \\ \midrule
NIST & 0.765 & 0.525 & 0.495 & 0.495 & 0.995 & 0.805 & 0.995 & \underline{0.995} & 0.525 & 0.5 & 0.945 & 0.96 & 0.66 & 0.97 \\
$\chi^2$ & 0.749 & 0.542 & 0.523 & 0.543 & 0.843 & 0.834 & 0.835 & 0.771 & 0.528 & 0.52 & 0.847 & 0.847 & 0.725 & 0.725 \\
HEDGE & 0.969 & 0.659 & 0.69 & 0.662 & 0.984 & 0.981 & 0.984 & 0.926 & 0.541 & 0.538 & 0.983 & 0.983 & 0.89 & 0.969 \\
EnCod & 0.978 & 0.86 & 0.74 & 0.852 & \underline{1.0} & \underline{1.0} & 0.998 & 0.9 & 0.65 & \underline{0.65} & \underline{1.0} & 0.999 & 0.97 & 0.925 \\\midrule
\fullname & \underline{0.996} & \underline{0.897} & \underline{0.809} & \underline{0.906} & 0.999 & \underline{1.0} & \underline{1.0} & 0.956 & \underline{0.686} & 0.634 & \underline{1.0} & \underline{1.0} & \underline{0.98} & \underline{0.984} \\
\bottomrule
\end{tabular}
}
\end{table}

\subsubsection{Comparative Analysis}
We compare \fullname to baseline approaches across varying fragment sizes. In line with~\cite{de2022reliable}, we first evaluate dedicated per-format binary classifiers for both \fullname and EnCoD, reporting per-format accuracy at 2KB fragment size. This setting assumes prior knowledge of the compressed format and establishes an upper bound on binary classification performance for ML-based approaches. Second, we evaluate aggregate binary classification performance across all fragment sizes. For \fullname and EnCoD, we use the multiclass classifier and collapse all compressed file-type predictions into a single ``compressed'' meta-class. This reflects a realistic operational setting where no prior information about the compressed format is known.
\begin{figure}[t]
  \begin{minipage}[t]{0.48\columnwidth}
    \includegraphics[width=\textwidth]{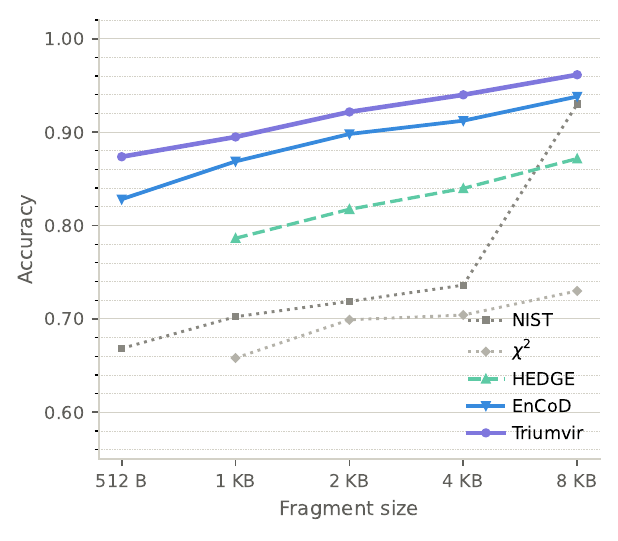}
    \caption{Aggregated binary accuracy.}
    \label{fig:binary_aggregate}
  \end{minipage}
  \hfill
  \begin{minipage}[t]{0.48\columnwidth}
    \includegraphics[width=\textwidth]{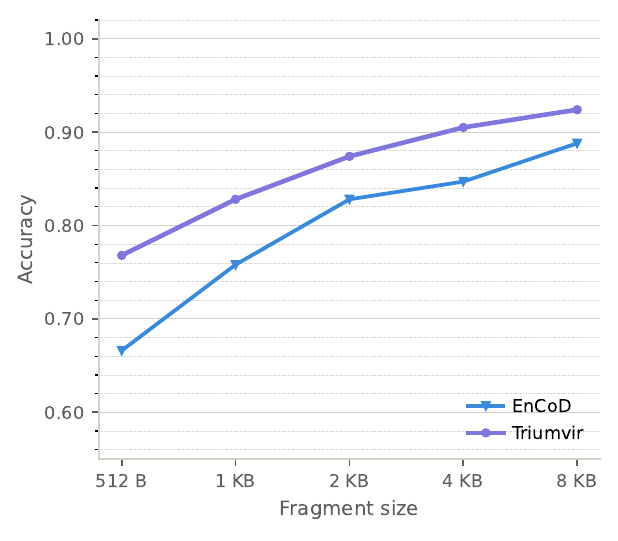}
    \caption{Multiclass Accuracy.}
    \label{fig:multiclass_aggregate}
  \end{minipage}
\end{figure}

\paragraph{Per-Format Performance}
Table~\ref{tab:2k_binary} reports binary classification performance at 2KB fragment size across all considered formats, highlighting clear differences in classification robustness across approaches at small fragment sizes. Classical statistical approaches perform poorly across most formats. NIST and $\chi^2$ achieve acceptable results only on classes with strong structural signals, such as images, mp3, and mpeg2/4. On generic compressed archives and h264/5 videos they severely underperform, approaching random-guessing performance. HEDGE improves over NIST and $\chi^2$ on most formats but remains well below ML-based approaches. EnCoD achieves strong results on structured formats, reaching perfect or near-perfect performance on image, audio, and mpeg/vp8 video formats. However, it struggles on generic compression and modern video codecs, where it achieves $0.74$ accuracy on rar and $0.65$ on h264/5.

\fullname achieves the best result on 11 out of 14 formats, matching EnCoD within $0.1$ percentage point (\textit{pp}) on png and generally improving over all baselines. On average, \fullname outperforms EnCoD by $4.45pp$ across generic compression formats, with the largest gain on rar ($+6.9pp$), gzip ($+5.4pp$), and zip ($+3.7pp$), suggesting that sequential and spatial experts capture structural patterns that are invisible to histogram-based representations. The most challenging formats for \fullname are h264 and h265 video formats, both below $0.69$ accuracy, consistent with the high compression efficiency of modern video codecs. H.265 is the only format where \fullname underperforms EnCoD ($-1.6pp$), representing a shared failure mode for all ML-based approaches at this fragment size.

\paragraph{Aggregated Binary Performance.}
Figure~\ref{fig:binary_aggregate} reports aggregate classification accuracy for \fullname and baselines across all fragment sizes. Statistical approaches generally perform poorly throughout, with the sole exception of NIST matching \encod at 8KB fragment sizes. HEDGE performs fairly well at larger sizes, reaching $0.84$ at 4KB and $0.87$ at 8KB, but lags significantly at lower fragment sizes. Both \fullname and \encod increase linearly with the squared of the fragment size, maintaining a consistent gap over statistical baselines across all fragment sizes. \fullname outperforms all baselines at every fragment size, with the most pronounced improvement at 512B ($+4.5pp$) and narrowing to $+2.3pp$ at 8KB. These results confirm that the multi-modal signal decomposition of \fullname provides consistent discriminative information, especially at lower fragment sizes.

As statistical approaches are shown to be consistently and substantially outperformed by ML-based approaches, subsequent analyses focus exclusively \fullname and EnCoD.

\begin{table}[t]
\centering
\caption{Detailed binary performance comparison between \fullname and EnCoD.}
\label{tab:binary_detailed}
\resizebox{\textwidth}{!}{%
\begin{tabular}{c | l|ccccccccccccccc}
\toprule
\textbf{Size} & \textbf{Method} & \textbf{bz2} & \textbf{zip} & \textbf{rar} & \textbf{gzip} & \textbf{png} & \textbf{jpeg} & \textbf{mp3} & \textbf{pdf} & \textbf{h264} & \textbf{h265} & \textbf{mpeg2} & \textbf{mpeg4} & \textbf{vp8} & \textbf{office} \\ 
\toprule

\multirow{2}{*}{\centering 512} 
& EnCoD  & 0.879 & 0.706 & 0.628 & 0.699 & \underline{1.0} & 0.975 & 0.96 & 0.81 & \underline{0.66} & \underline{0.66} & 0.992 & 0.99 & 0.92 & 0.766 \\
& \fullname & \underline{0.933} & \underline{0.74} & \underline{0.668} & \underline{0.754} & 0.994 & \underline{0.984} & \underline{0.993} & \underline{0.863} & 0.597 & 0.575 & \underline{1.0} & \underline{0.997} & \underline{0.943} & \underline{0.926} \\ 
\midrule

\multirow{2}{*}{\centering 1K} 
& EnCoD  & 0.94 & 0.785 & 0.681 & 0.777 & \underline{1.0} & 0.996 & 0.991 & 0.87 & \underline{0.660} & \underline{0.661} & 0.998 & 0.998 & 0.94 & 0.843 \\
& \fullname & \underline{0.974} & \underline{0.84} & \underline{0.732} & \underline{0.811} & 0.999 & \underline{0.999} & \underline{1.0} & \underline{0.903} & 0.622 & 0.622 & \underline{1.0} & \underline{1.0} & \underline{0.954} & \underline{0.96} \\ 
\midrule

\multirow{2}{*}{\centering 2K} 
& EnCoD  & 0.978 & 0.86 & 0.74 & 0.852 & \underline{1.0} & \underline{1.0} & 0.998 & 0.9 & 0.65 & 0.650 & \underline{1.0} & 0.999 & 0.97 & 0.925 \\
& \fullname & \underline{0.996} & \underline{0.897} & \underline{0.809} & \underline{0.906} & 0.999 & \underline{1.0} & \underline{1.0} & \underline{0.956} & \underline{0.686} & \underline{0.662} & \underline{1.0} & \underline{1.0} & \underline{0.98} & \underline{0.984} \\ 
\midrule

\multirow{2}{*}{\centering 4K} 
& EnCoD  & 0.99 & 0.926 & 0.81 & 0.92 & \underline{1.0} & \underline{1.0} & \underline{1.0} & \underline{0.97} & 0.68 & \underline{0.73} & \underline{1.0} & \underline{1.0} & 0.99 & 0.92 \\
& \fullname & \underline{1.0} & \underline{0.958} & \underline{0.885} & \underline{0.962} & \underline{1.0} & \underline{1.0} & \underline{1.0} & 0.969 & \underline{0.757} & 0.705 & 0.999 & \underline{1.0} & \underline{0.994} & \underline{0.992} \\ 
\midrule

\multirow{2}{*}{\centering 8K} 
& EnCoD  & 0.998 & 0.975 & 0.88 & 0.968 & \underline{1.0} & \underline{1.0} & \underline{1.0} & \underline{0.99} & 0.77 & 0.76 & \underline{1.0} & \underline{1.0} & \underline{1.0} & 0.953 \\
& \fullname & \underline{0.999} & \underline{0.995} & \underline{0.938} & \underline{0.992} & \underline{1.0} & \underline{1.0} & \underline{1.0} & 0.978 & \underline{0.825} & \underline{0.791} & \underline{1.0} & \underline{1.0} & 0.999 & \underline{0.996} \\
\bottomrule
\end{tabular}%
}
\end{table}

\subsubsection{In-depth Comparison}
This section provides an in-depth comparison of binary classification performance for \fullname and EnCoD. Table~\ref{tab:binary_detailed} reports per-format binary classification accuracy across all fragment sizes. \fullname consistently outperforms EnCoD on generic compression formats across all fragment sizes. The improvements are most pronounced on rar archives, where \fullname leads by $4pp$ at 512B, $5.1pp$ at 1KB, $6.9pp$ at 2KB, $7.5pp$ at 4KB, and $5.8pp$ at 8KB. On average, on generic compressed formats \fullname outperforms EnCoD by $4.45pp$ on sub 4KB sizes and by $3.98pp$ overall. These results confirm our hypothesis that back-references and repetition coding structures typical of compression cannot be fully captured by the statistical expert alone, and validate the contribution of the sequential and spatial experts.

On png, jpeg, mp3, and mpeg structured formants, both approaches quickly converge to near-perfect accuracy, with accuracy $\geq0.99$ at 1KB. While the statistical expert is mostly sufficient in this regime, it is worth noting that at 512B size spatial and statistical signals still provide measurable improvements ($+3.3pp$ on mp3). Furthermore, on pdf files \fullname significantly outperforms \encod, epsecially at 512B size ($+5.3pp$). These results suggest that spatial and structural signals provide meaningful discriminative value at the smallest fragment sizes, where byte-frequency distribution is least reliable.

The most challenging formats for both approaches are h264 and h265, which represent the only consistent failure mode across all fragment sizes. At 512B, \fullname underperforms \encod on both formats; a notable reversal of the general trend. This suggest that, at small fragment sizes, the sequential and spatial experts introduce noise rather than signal. This is possibly because h264 and h265 codec outputs lack significant structural signal at small fragment sizes. As fragment size increases, \fullname recovers and significantly outperforms \encod on h264 from 4KB onward ($+7.7pp$ at 4KB, $+5.5pp$ at 8KB) and is comparable or outperforms on h265. However, h265 remains the single most challenging format across al sizes, never exceeding $0.791$ accuracy.

Office documents represent the largest absolute improvement of \fullname over \encod, with a gain of $+16.0pp$ at 512B that narrows progressively to $+4.3pp$ at 8KB. This result is particularly noteworthy, as Office documents utilize DEFLATE compression internally, the same algorithm underlying zip and gzip. Yet, \fullname performs substantially better on Office than either zip or gzip ($0.926$ vs $0.740$ and $0.754$ at 512B, respectively). This discrepancy suggests that the richer internal structure of Office files produces byte structures that are well-suited for the sequential and spatial experts, providing further evidence towards our hypothesis.

\subsection{Multiclass Classification Analysis}
\label{sec:multiclass_analysis}
\begin{table}[t]
\centering
\scriptsize
\caption{Multiclass Detailed Performance Comparison}
\label{tab:multiclass_detail}
\begin{tabular}{l|l|ccccc}
\toprule
\textbf{Metric} & \textbf{Model} & \textbf{512} & \textbf{1K} & \textbf{2K} & \textbf{4K} & \textbf{8K} \\
\midrule
\multirow{2}{*}{Accuracy}
& EnCoD     & 0.666 & 0.758 & 0.828 & 0.847 & 0.888 \\
& \fullname & \underline{0.730} & \underline{0.802} & \underline{0.847} & \underline{0.885} & \underline{0.916} \\
\midrule
\multirow{2}{*}{Precision}
& EnCoD     & 0.704 & 0.797 & 0.828 & 0.858 & 0.892 \\
& \fullname & \underline{0.757} & \underline{0.820} & \underline{0.859} & \underline{0.890} & \underline{0.919} \\
\midrule
\multirow{2}{*}{Recall}
& EnCoD     & 0.666 & 0.758 & 0.828 & 0.847 & 0.888 \\
& \fullname & \underline{0.730} & \underline{0.802} & \underline{0.847} & \underline{0.885} & \underline{0.916} \\
\midrule
\multirow{2}{*}{F1-score}
& EnCoD     & 0.685 & 0.777 & 0.828 & 0.852 & 0.890 \\
& \fullname & \underline{0.731} & \underline{0.802} & \underline{0.846} & \underline{0.885} & \underline{0.916} \\
\bottomrule
\end{tabular}
\end{table}

\subsubsection{Comparative Analysis}
In line with~\cite{de2022reliable}, we evaluate multiclass classification performance by aggregating file types in 8 macro-classes: encrypted (ENC), general compression (CMP), audio (MP3), PNG, JPG, office (OFF), PDF, and video formats (VIDEO). Table~\ref{tab:multiclass_detail} reports a consistent improvement of \fullname over EnCoD across all chunk sizes and evaluation metrics. The gains are particularly pronounced at smaller fragment sizes, where the classification task is inherently more challenging due to limited observable structure.
At 512B, \fullname achieves an accuracy of $0.73$ compared to $0.666$ for EnCoD, corresponding to a $+6.4pp$ improvement. Similar improvements are observed for F1-score ($+4.5pp$) and precision ($+5.3pp$). This trend indicates that \fullname is substantially more robust in low-information regimes, where histogram-based representations alone are insufficient. At 1KB, the improvement remains significant, with accuracy increasing from $0.758$ to $0.80$ ($+4.4pp$), and precision from $0.797$ to $0.82$ ($+2.3pp$), confirming that the benefits of the proposed approach persist as more context becomes available. As fragment size increases, the performance gap narrows but remains consistent. At 2KB, \fullname improves accuracy by $+1.9pp$, while at 4KB the gain is $+3.8pp$. At 8KB, where both methods approach saturation, \fullname still maintains a measurable advantage ($+2.8pp$ in accuracy and $+2.6pp$ in F1-score). Notably, precision improvements remain stable across sizes, suggesting that \fullname reduces false positives more effectively.

Overall, these results highlight two key insights. First, multi-modal modeling provides the largest benefit when data is scarce, effectively compensating for the lack of strong statistical signals. Second, even at larger sizes where distributional features become more informative, \fullname continues to exploit complementary structural cues, leading to consistent, albeit smaller, performance gains.
\vspace{-1em}
\subsubsection{Error Analysis}
\begin{figure}[t]
  \begin{subfigure}[t]{0.48\columnwidth}
    \includegraphics[width=\textwidth]{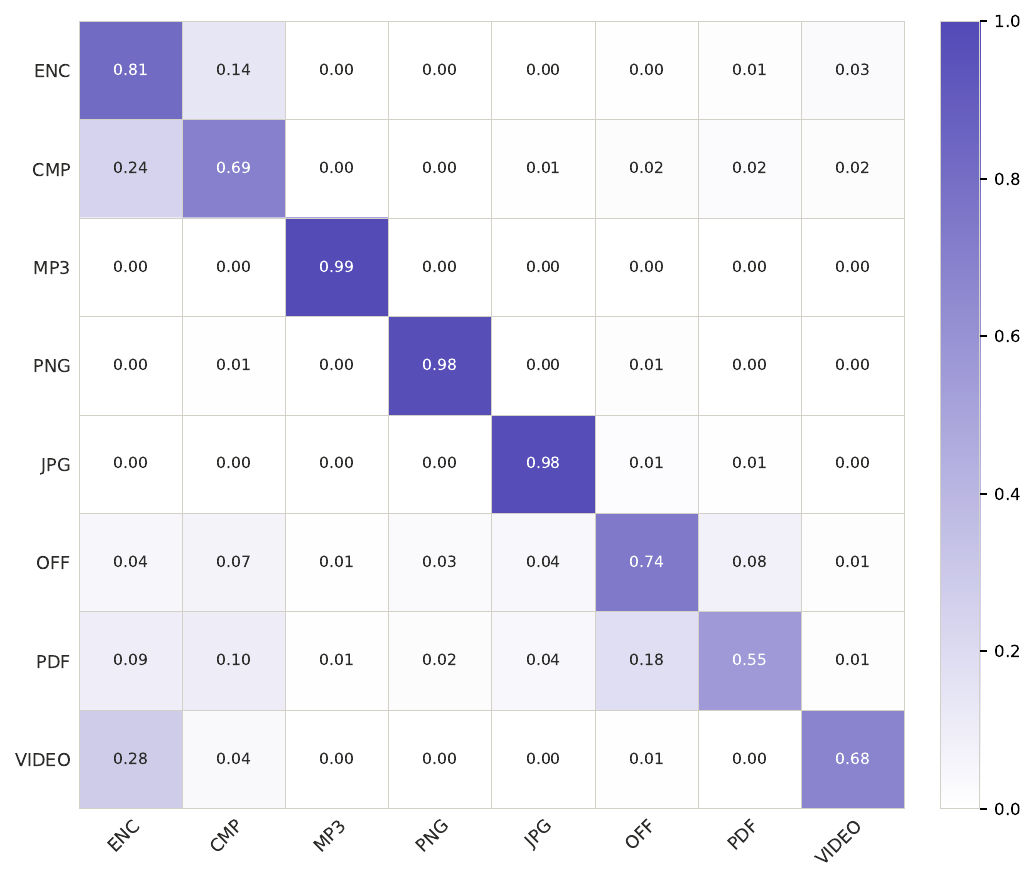}
    \caption{1KB fragment size.}
    \label{fig:conf1k}
  \end{subfigure}
  \hfill
  \begin{subfigure}[t]{0.48\columnwidth}
    \includegraphics[width=\textwidth]{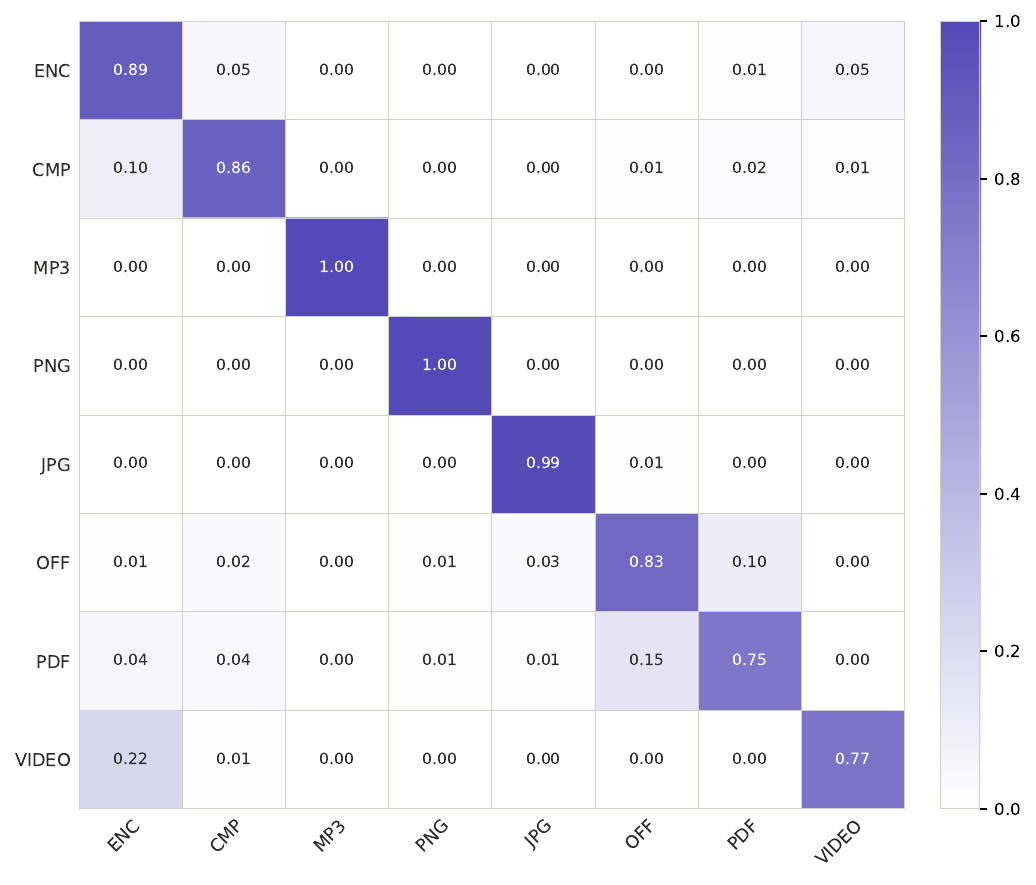}
    \caption{4KB fragment size.}
    \label{fig:conf4k}
  \end{subfigure}
  \caption{Multiclass classification confusion matrix for \fullname.}
  \label{fig:conf_mat}
\end{figure}

Figures~\ref{fig:conf1k} and~\ref{fig:conf4k} present the confusion matrices for \fullname at 1KB and 4KB fragment sizes. The error structure is consistent across both sizes, with misclassifications concentrated mainly in three patterns: ENC-CMP, PDF-OFF, and VIDEO-ENC. Fragment size affects only the magnitude of the misclassificaiton, rather than their nature. This indicates that the errors derive from fundamental overlaps in the signal of the classes, rather than size-related artifacts. The dominant error pattern is the mutual misclassification between encrypted and generic compressed fragments. At 1KB, $14\%$ of encrypted fragments are misclassified as compressed, and $24\%$ of compressed are misclassified as encrypted. Both figures improve substantially at 4KB size ($6, 10\%$), with the misclassification following the same pattern. We note that the misclassification rate of compressed to encrypted is consistently higher than encrypted to compressed. We speculate that this may be due to the low intra-class variance of encrypted fragments compared to the heterogeneous compressed class, which leads to a well-defined region for the encrypted class in the feature space, compared to a more sparse region for the higher-variance compressed class.

Regarding the other two primary error patterns, VIDEO fragments are consistently misclassified as encrypted at both 1KB and 4KB. This is the most persistent error in the matrix, and is in line with the binary results in Section~\ref{sec:binary_analysis}. Nonetheless, we highlight that on the VIDEO class \fullname at 1KB size outperforms \encod at 2KB size, showcasing the benefits of our approach. Pdf fragments consistently show the weakest recall among all classes, with persistent mutual confusion with the Office class. The distributed misclassification pattern of pdf is consistent with its heterogeneous internal structure, which embeds multiple different content types. Further analysis on this error mode appears a good direction for future works.

\subsection{Ablation Study}
\label{sec:ablation_analysis}

Table~\ref{tab:ablation} presents an ablation analysis at 2KB fragment size, evaluating the contribution of the different signal spaces captured by the three experts. The results show clear performance difference across pairwise combinations highlighting the complementary nature of the three feature representations. Among the combination pairs, \textit{Sequential + Spatial} yields the lowest performance with an accuracy of $0.797$, indicating that without statistical features, the models struggle to capture sufficient discriminative signals. Leveraging statistical information leads to substantial improvements. The \textit{Spatial + Statistical} combination increases accuracy to $0.83$ ($+3.3pp$) over the previous combination, while the \textit{Sequential + Statistical} further improves to $0.839$ ($+4.2pp$ and $+0.9pp$) over the previous two combinations. This suggests that the statistical features provide a strong baseline signal, and that sequential features contribute more effectively than the spatial ones when combined with the statistical model. This trend is present across all performance metrics, confirming the consistency of such effects. Comparing these variants against the full \fullname architecture, we observe further improvements across all metrics. Specifically, \fullname achieves an accuracy of $0.847$, outperforming the best two-modal combination (\textit{Sequential + Statistical}) by $+0.74pp$ and the weakest combination by nearly $+5pp$. Precision follows a similar pattern, with \fullname reaching $0.859$, exceeding the strongest baseline by approximately $+0.8pp$. Gains in F1-score also follow a similar trend.

These results demonstrate that each signal space captures unique discriminative features that positively contribute to final classification performance. The gated architecture in \fullname effectively leverages this complementarity by adaptively conditioning expert outputs based on an explicit per-sample confidence signal, leading to consistently superior performance over any fixed combination.

\subsection{Overhead}
\label{sec:eval_overhead}

\begin{table}[t]
\centering
\scriptsize
\caption{Ablation performance comparison on multiclass classification, 2KB size.}
\label{tab:ablation}
\begin{tabular}{l|cccc}
\toprule
\textbf{Model} & \textbf{Accuracy} & \textbf{Precision} & \textbf{Recall} & \textbf{F1-score} \\ 
\midrule
Sequential + Spatial        & 0.7971 & 0.8134 & 0.7971 & 0.7949 \\
Spatial + Statistical       & 0.8304 & 0.8493 & 0.8304 & 0.8312 \\
Sequential + Statistical    & 0.8396 & 0.8506 & 0.8396 & 0.8397 \\ \midrule
\fullname & \underline{0.847} & \underline{0.859} & \underline{0.847}  & \underline{0.846} \\
\bottomrule
\end{tabular}
\end{table}

\begin{table}[t]
\scriptsize
\centering
\caption{Time required by \fullname to classify one sample, in seconds.}
\begin{tabular}{l|c c c c c} \hline
 \textbf{Size} & \textbf{512B} & \textbf{1K} & \textbf{2K} & \textbf{4K} & \textbf{8K} \\\hline
 \textbf{Average} & 0.0157 & 0.0327 & 0.0880 & 0.2148 & 0.6942 \\
 \textbf{Std.dev} & 0.0007 & 0.0007 & 0.0012 & 0.0011 & 0.0051 \\\hline
\end{tabular}
\label{tab:overhead}
\end{table}
In the final part of our evaluation, we analyze the practical applicability of \fullname, comparing its runtime in order to understand if it can be deployed in time-critical applications. For this test, we used a small dataset comprised of 1000 randomly-selected compressed or encrypted samples for each of the considered chunk sizes. 
We ran \fullname on each sample, taking  individual runtime. Table~\ref{tab:overhead} presents the results of our evaluation.  We notice a monotonic increase in classification time as the input size grows from 512B to 8K. Both the average runtime and its standard deviation scale upward with input size. Importantly, despite this growth trend, the absolute prediction times remain quite reasonable for most chunk sizes. Up to 2K, the average latency stays low (sub-0.1 seconds), which suggests that the system can handle typical small-to-medium inputs with negligible overhead in practice. Even at 4K, the runtime remains around 0.21 seconds, which is still acceptable for many real-time or near-real-time applications. Only at the largest tested size (8K) does the cost approach a level that may become noticeable in latency-sensitive settings. However, \fullname can easily classify multiple samples in parallel by exploiting the heavy parallelism of GPUs, further decreasing the runtime required per individual sample.
\section{Conclusion}\label{sec:conclusion}
Automated analysis of raw byte content and the ability to characterize the file type of such content is important for a variety of security applications. Specifically, distinguishing between encrypted and compressed content remains critical, with direct implication in tasks such as ransomware detection, memory forensics and more.  
This paper pushes the state-of-the-art in this area by introducing \fullname, a multi-modal, uncertainty-aware ensemble architecture designed to capture complementary views of data fragments, including both distributional and sequential characteristics. Extensive experimental analysis showed that \textit{\fullname} consistently outperforms state-of-the-art methods with gains up to $+4.5pp$ in binary and $+6.4pp$ in multiclass classification. Ablation studies confirm that combining modalities is critical, yielding improvements of up to $+5pp$ over partial configurations.

\bibliographystyle{unsrt}
\bibliography{bibliography}

\appendix
\section{Appendix}

\subsection{Additional Evaluation Details}
We provide further details on the experimental comparison between \fullname and \encod in the multiclass setting. Table~\ref{tab:multiclass_appendix} presents per-expert performance, \fullname performance, and \encod performance across fragment sizes.

\begin{table}[t]
\centering
\scriptsize
\caption{Detailed multiclass performance comparison}
\label{tab:multiclass_appendix}
\begin{tabular}{lccccc}
\toprule
\textbf{Model} & \textbf{Size} & \textbf{Accuracy} & \textbf{Precision} & \textbf{Recall} & \textbf{F1-score} \\ \midrule
Statistical              & 512B  & 0.670 & 0.707 & 0.671 & 0.674 \\
Spatial              & 512B  & 0.548 & 0.604 & 0.548 & 0.550 \\
Sequential         & 512B  & 0.618 & 0.630 & 0.618 & 0.605 \\  \midrule
\fullname           & 512B  & 0.730 & 0.757 & 0.730 & 0.731 \\
EnCoD               & 512B  & 0.666 & 0.704 & 0.666 & 0.685 \\
\midrule
Statistical      & 1024B & 0.758 & 0.788 & 0.758 & 0.760 \\
Spatial          & 1024B & 0.670 & 0.720 & 0.670 & 0.673 \\
Sequential          & 1024B & 0.715 & 0.729 & 0.715 & 0.709 \\ \midrule
\fullname           & 1024B & 0.802 & 0.820 & 0.802 & 0.802 \\
EnCoD               & 1024B & 0.758 & 0.797 & 0.758 & 0.777 \\
\midrule
Statistical      & 2048B  & 0.818 & 0.835 & 0.818 & 0.818 \\
Spatial          & 2048B  & 0.738 & 0.775 & 0.738 & 0.741 \\
Sequential          & 2048B  & 0.755 & 0.762 & 0.755 & 0.749 \\ \midrule
\fullname           & 2048B  & 0.847 & 0.859 & 0.847 & 0.846 \\
EnCoD               & 2048B  & 0.828 & 0.828 & 0.828 & 0.828 \\
\midrule
Statistical      & 4096B & 0.861 & 0.872 & 0.861 & 0.862 \\
Spatial          & 4096B & 0.794 & 0.821 & 0.794 & 0.795 \\
Sequential          & 4096B & 0.768 & 0.778 & 0.768 & 0.761 \\ \midrule
\fullname           & 4096B & 0.885 & 0.890 & 0.885 & 0.885 \\
EnCoD               & 4096B & 0.847 & 0.858 & 0.847 & 0.852 \\
\midrule
Statistical      & 8192B  & 0.900 & 0.903 & 0.900 & 0.900 \\
Spatial          & 8192B  & 0.841 & 0.850 & 0.841 & 0.841 \\ 
Sequential          & 8192B  & 0.761 & 0.760 & 0.761 & 0.756 \\ \midrule
\fullname           & 8192B  & 0.916 & 0.919 & 0.916 & 0.916 \\
EnCoD               & 8192B & 0.888 & 0.892 & 0.888 & 0.89 \\
\bottomrule

\end{tabular}%
\end{table}

\subsection{Dataset}
\label{sec:dataset}
For a fair comparison with the baseline approaches, we use the Encod~\cite{de2020encod} dataset with fragment sizes of 512, 1024, 2048, 4096, and 8192 bytes. The dataset comprises 400 million encrypted and compressed fragments spanning 17 different data formats, covering all major content types and encrypted data:

\begin{enumerate}

\item \textbf{Encrypted data (\textit{enc})}. Encrypted data fragments were generated using the AES-256-CBC through the PyCryptodome library~\footnote{https://pycryptodome.readthedocs.io}.

\item \textbf{Compressed archives (zip, gzip, rar, bz2):} covering
DEFLATE, rar, and Burrows–Wheeler compression. These are among the most widely used algorithms for general-purpose file compression and network protocols.

\item \textbf{Images (png, jpeg):} png uses lossless DEFLATE compression with a file structure distinct from zip archives. jpeg uses DCT-based lossy compression.

\item \textbf{Audio (mp3):} mp3 uses a psychoacoustic model to remove perceptually inaudible frequencies, followed by lossy compression based on a modified DCT transform.

\item \textbf{Documents (pdf):} pdf files internally consist of a tree of objects that may be compressed using various techniques. Most pdf documents contain substantial compressed content including embedded images.

\item \textbf{Microsoft Office documents:} Office file formats use DEFLATE compression internally across all major document types.

\item \textbf{Video (h264, h265, mpeg2, mpeg4, vp8):} A representative set of video codecs spanning current standards and legacy formats.
\end{enumerate}

\end{document}